\documentclass[10pt,twocolumn,letterpaper]{article}

\usepackage[pagenumbers]{iccv} 
\usepackage{pifont}
\newcommand{\cmark}{\ding{51}}%
\newcommand{\xmark}{\ding{55}}%
\usepackage{subcaption}

\definecolor{iccvblue}{rgb}{0.21,0.49,0.74}
\usepackage[pagebackref,breaklinks,colorlinks,allcolors=iccvblue]{hyperref}

\def\paperID{11} 
\def\confName{ICCV CV4DC}
\def\confYear{2025}

\title{MSVD-Indonesian: A Benchmark for Multimodal Video-Text Tasks in Indonesian}

\author{
  Willy Fitra Hendria \\
  Independent Researcher \\
  Seoul, South Korea\\
  {\tt\small willyfitrahendria@gmail.com} \\
 \\
}

\begin{document}
\maketitle
\begin{abstract}
Multimodal learning on video and text has seen significant progress, particularly in tasks like text-to-video retrieval, video-to-text retrieval, and video captioning. However, most existing methods and datasets focus exclusively on English. Despite Indonesian being one of the most widely spoken languages, multimodal research in Indonesian remains under-explored, largely due to the lack of benchmark datasets. To address this gap, we introduce the first public Indonesian video-text dataset by translating the English captions in the MSVD dataset into Indonesian. Using this dataset, we evaluate neural network models which were developed for the English video-text dataset on three tasks, i.e., text-to-video retrieval, video-to-text retrieval, and video captioning. Most existing models rely on feature extractors pretrained on English vision-language datasets, raising concerns about their applicability to Indonesian, given the scarcity of large-scale pretraining resources in the language. We apply a cross-lingual transfer learning approach by leveraging English-pretrained extractors and fine-tuning models on our Indonesian dataset. Experimental results demonstrate that this strategy improves performance across all tasks and metrics. We release our dataset publicly to support future research and hope it will inspire further progress in Indonesian multimodal learning\footnote{\label{fn:github}\href{https://github.com/willyfh/msvd-indonesian}{https://github.com/willyfh/msvd-indonesian}}.
\end{abstract}    
\section{Introduction}
\label{sec:intro}

Multimodal machine learning \cite{10.1109/TPAMI.2018.2798607} enables models to learn from multiple modalities such as text, vision, and audio. Recent advances in this field have led to progress in video-text tasks, including text-to-video retrieval \cite{DBLP:journals/corr/TorabiTS16}, video-to-text retrieval \cite{PerezMartin2021ACR}, and video captioning \cite{venugopalan15iccv}. These tasks typically rely on supervised training over large-scale datasets consisting of paired video and textual descriptions.

However, the majority of available video-text dataset, e.g., MSVD \cite{chen:acl11}, MSR-VTT \cite{xu2016msr-vtt}, and ActivityNet Captions \cite{krishna2017dense}, are constructed in English. Only a few multilingual datasets exist for languages such as Chinese \cite{Wang_2019_ICCV}, Turkish \cite{8806555}, Italian \cite{IJCOL:scaiella_et_al:2019}, and Hindi \cite{10.1007/s00530-021-00816-3}. Despite Indonesian being one of the most spoken languages globally, there is currently no publicly available video-text dataset for it. This lack of resources limits the development and evaluation of multimodal systems for Indonesian.

To address this gap, we construct the first public Indonesian video-text dataset by translating English captions in the MSVD dataset into Indonesian. Our MSVD-Indonesian dataset consists of 1,970 videos and approximately 80,000 Indonesian sentences, mirroring the structure and scale of the original dataset.

We use our dataset to evaluate neural network models originally developed for English video-text tasks. Specifically, we adopt X-CLIP \cite{10.1145/3503161.3547910} for retrieval and VNS-GRU \cite{DBLP:conf/ecai/Chen0020a} for captioning. These models rely on pretrained feature extractors like CLIP \cite{Radford2021LearningTV} or semantic concept detection (SCD) \cite{SCN_CVPR2017}, which are trained on predominantly English data. To adapt them for Indonesian, we apply cross-lingual transfer learning by reusing English-pretrained extractors and fine-tuning them on our dataset.

Our experiments demonstrate that this transfer learning approach improves performance across all three tasks. We release the MSVD-Indonesian dataset publicly and hope it encourages further research in multilingual and low-resource video-text learning.

In summary, our main contributions are listed as follows:
\begin{itemize}
    \item We release the first Indonesian video-text dataset, translated from MSVD.
    \item We establish baseline results for three tasks using models originally developed for English video-text tasks.
    \item We demonstrate that cross-lingual transfer learning is effective for the Indonesian video-text tasks.
    \item We outline future directions enabled by this dataset.
\end{itemize}

\section{Related Work}
\label{sec:related}

\subsection{Video-Text Datasets}
Video-text datasets support a variety of multimodal tasks, such as video captioning \cite{venugopalan15iccv} and retrieval \cite{10.1145/3503161.3547910}. Most existing datasets, including MSVD \cite{chen:acl11}, MSR-VTT \cite{xu2016msr-vtt}, and ActivityNet Captions \cite{krishna2017dense}, are constructed with English annotations. A few multilingual datasets exist, including MSVD-CN \cite{msvdcn}, MSVD-Turkish \cite{8806555}, and others in Chinese, Hindi, and Italian \cite{Wang_2019_ICCV, IJCOL:scaiella_et_al:2019, 10.1007/s00530-021-00816-3}. Although the original MSVD dataset was collected in multiple languages, the publicly released version by Chen and Dolan \cite{chen:acl11} includes only English. Later efforts reconstructed Chinese and Turkish versions from the English set. To our knowledge, no Indonesian version exists. Our work addresses this by constructing and releasing MSVD-Indonesian, the first public video-text dataset in Indonesian.

\subsection{Video-Text Retrieval}
Video-text retrieval encompasses text-to-video and video-to-text retrieval tasks. Recent methods have benefited significantly from vision-language pretraining. Luo \etal proposed CLIP4Clip \cite{10.1145/3474085.3479207}, which uses CLIP \cite{Radford2021LearningTV} to encode video frames and text into a shared space. Ma \etal extended this with X-CLIP \cite{10.1145/3503161.3547910}, introducing multi-granular contrastive learning to improve retrieval performance. While these methods perform well on English benchmarks, their cross-lingual effectiveness remains untested. In our work, we apply X-CLIP to our Indonesian dataset using cross-lingual transfer learning and assess its effectiveness on both retrieval tasks.

\subsection{Video Captioning}
In video captioning, many models use pretrained semantic extractors. One such model is the Semantic Concept Detector (SCD) by Gan \etal \cite{SCN_CVPR2017}, which predicts keywords from videos in a multi-label fashion. SCD-based approaches have been incorporated into various captioning models \cite{DBLP:conf/ecai/Chen0020a, 10.3389/frobt.2020.475767, Perez-Martin_2021_WACV, 9412898}. Among them, VNS-GRU \cite{DBLP:conf/ecai/Chen0020a} achieves strong results on MSVD. In this work, we adopt VNS-GRU and adapt it to our Indonesian dataset using SCD pretrained on English annotations. This setup allows us to investigate whether semantic features learned in English can transfer to a low-resource language scenario.

\section{MSVD-Indonesian Dataset}

\subsection{Dataset Collection}
The MSVD dataset originally contained 2089 videos \cite{chen:acl11}, but due to removed YouTube links, only 1970 videos were retained and widely adopted in prior research. We use this reconstructed version, including 80,827 English sentences, and translate all sentences to Indonesian using the Google Translate API. Each video in our dataset has the same number of sentences as in MSVD, preserving a one-to-one mapping. This alignment enables multilingual research and cross-lingual benchmarking.

\subsection{Dataset Analysis}
\begin{figure}[t!]
  \centering  
  \begin{subfigure}[b]{0.495\textwidth}
    \includegraphics[width=\textwidth]{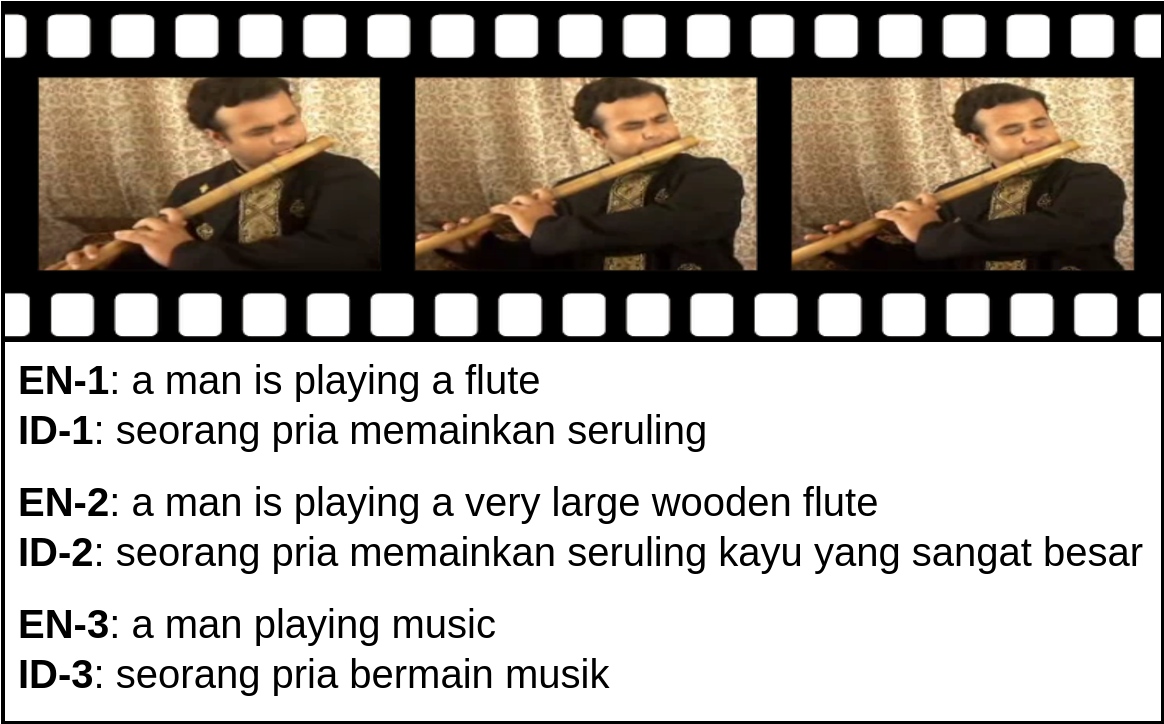}
    \caption{}
  \end{subfigure}
  \hfill
  \begin{subfigure}[b]{0.495\textwidth}
    \includegraphics[width=\textwidth]{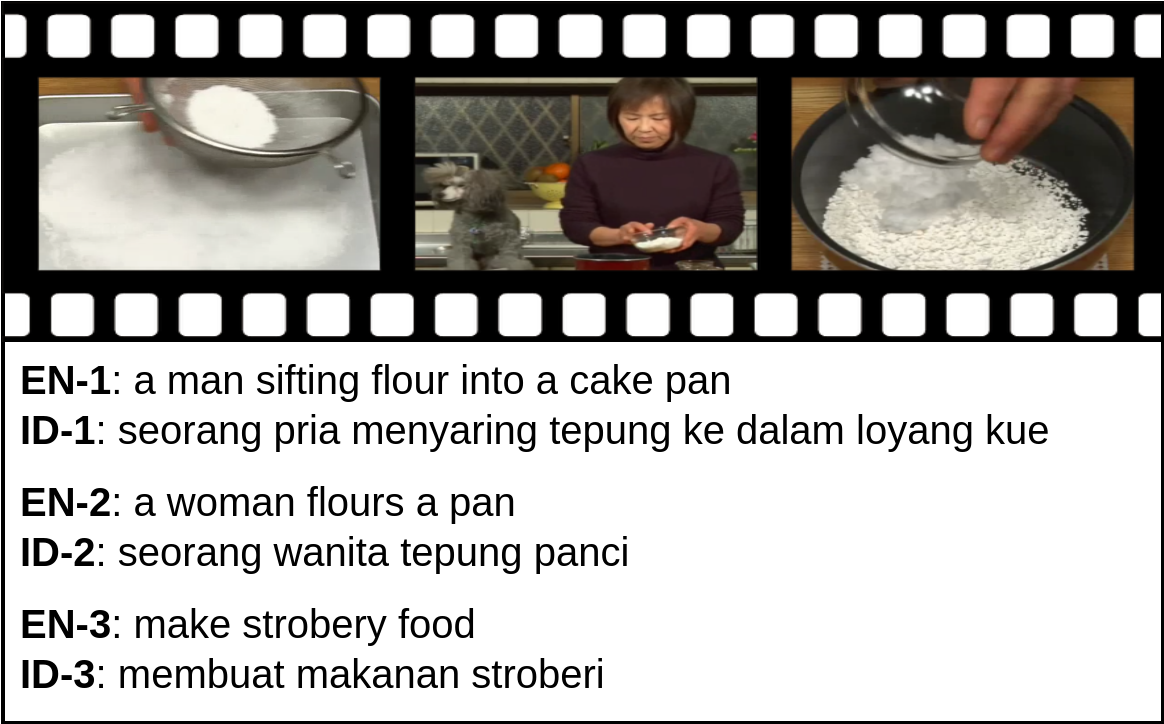}
    \caption{}
  \end{subfigure}
  \caption{Samples from the MSVD-Indonesian dataset. Each video is shown with three English (EN-\#) and Indonesian (ID-\#) sentence pairs.}
  \label{fig:sample-dataset}
\end{figure}

Machine-translated data can contain grammatical or semantic inaccuracies. While many Indonesian sentences are translated well, others contain artifacts. Figure~\ref{fig:sample-dataset}(a) shows samples where the translations closely match the English captions. Figure~\ref{fig:sample-dataset}(b) shows issues such as mistranslating the verb “flours” to “tepung.” In some cases, errors originate from the original English annotations, such as a mislabeled mention of "strobery" that doesn't appear in the video. Interestingly, the translation of “strobery” is corrected to the standard Indonesian “stroberi.” We retain such noisy samples to reflect real-world dataset imperfections.

\begin{figure*}[htbp]
  \centering  
  \begin{subfigure}[b]{0.45\textwidth}
    \includegraphics[width=\textwidth]{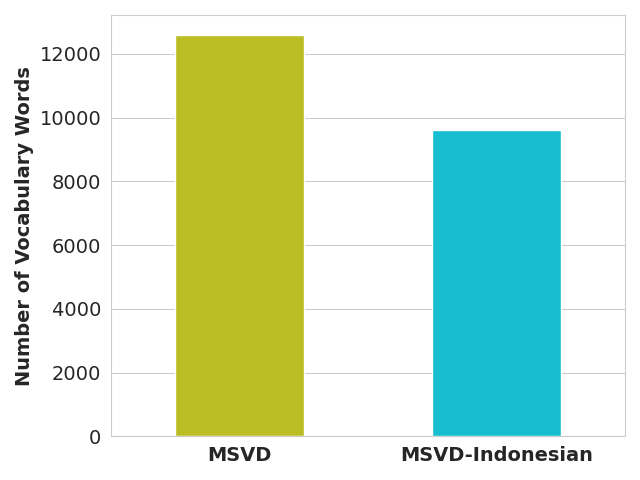}
    \caption{}
  \end{subfigure}
  \hfill
  \begin{subfigure}[b]{0.45\textwidth}
    \includegraphics[width=\textwidth]{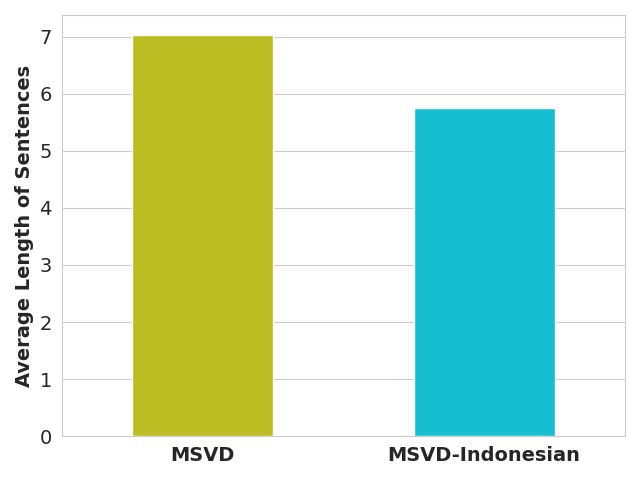}
    \caption{}
  \end{subfigure}
  \caption{Comparison of MSVD and MSVD-Indonesian datasets based on (a) vocabulary size and (b) average sentence length.}
  \label{fig:dataset-analysis}
\end{figure*}

We also analyze vocabulary size and sentence length. As shown in Figure~\ref{fig:dataset-analysis}, the MSVD-Indonesian dataset contains 9,457 unique words, compared to 12,592 in English. The average sentence length is 5.7 words in Indonesian and 7 words in English. These differences suggest that models trained on MSVD-English may not directly transfer well to Indonesian, highlighting the need for adaptation and re-evaluation on our dataset.

\section{Methods}

We adopt two models previously applied to English video-text tasks that rely on pretrained vision-language features: X-CLIP \cite{10.1145/3503161.3547910} for text-video retrieval, and VNS-GRU \cite{DBLP:conf/ecai/Chen0020a} for video captioning. Our setup assumes a low-resource scenario in which large-scale vision-language pretraining data for Indonesian is unavailable. Accordingly, we use CLIP \cite{Radford2021LearningTV} and SCD \cite{SCN_CVPR2017} pretrained on English data, and fine-tune both models on our MSVD-Indonesian dataset.

\subsection{X-CLIP}
X-CLIP extends CLIP4Clip \cite{10.1145/3474085.3479207} by introducing multi-granular contrastive learning across video-sentence, video-word, sentence-frame, and frame-word levels. We follow its standard setup and extract frame features from 1 FPS sampled frames using the CLIP visual encoder. Sentence- and word-level features are obtained from the text encoder. Unlike the original model, we omit the temporal encoder, as our experiments show it degrades performance on our dataset. Final video features are computed by averaging frame embeddings. Cross-modal similarities are fused using the attention-over-similarity module (AOSM) \cite{10.1145/3503161.3547910}. We refer readers to the original paper for architectural details.

\subsection{VNS-GRU}
VNS-GRU integrates semantic features from an SCD model \cite{SCN_CVPR2017}, which is pretrained to predict frequent keywords in a video. The output semantic vector is concatenated with features from an ImageNet-pretrained classifier. Additionally, video features are extracted using a 3D CNN pretrained on Kinetics-400 \cite{DBLP:journals/corr/KayCSZHVVGBNSZ17}. The final GRU-based decoder is trained using techniques such as variational dropout, layer normalization, and comprehensive selection, following \cite{DBLP:conf/ecai/Chen0020a}. We retain the original architecture and focus on evaluating its effectiveness in the Indonesian language setting.

\section{Experimental Results}
\label{sec:experiments}
In this section, we discuss our experiment details for the retrieval and captioning tasks, including the evaluation metrics and the implementation details. We then discuss our experimental results on the test set, which include quantitative and qualitative results. For both retrieval and captioning tasks, we follow the standard split of the MSVD dataset, i.e., 1200, 100, and 670 videos for train, validation, and test set.

\subsection{Evaluation Metrics}
\subsubsection{Retrieval}
We evaluate our retrieval experiments by using five commonly used metrics in text-to-video and video-to-text retrieval tasks, i.e., R@1, R@5, R@10, MedianRank, and MeanRank. 1) R@1, R@5, and R@10 (recall at $K$) measure the proportion of relevant items correctly retrieved among the top $K$ items. 2) Median rank indicates the median position at which the relevant items are found in a ranked list. 3) Mean rank, on the other hand, measure the average position of the relevant items found in the same list.

\subsubsection{Captioning}
For the captioning experiments, we assess the performance of the models by using four popular metrics in video captioning, i.e., BLEU@4 \cite{10.3115/1073083.1073135}. ROUGE-L \cite{lin2004rouge}, METEOR \cite{banarjee2005}, and CIDEr \cite{7299087}. 1) BLEU@4 computes the accuracy of a method by taking the precision of the generated sentences in terms of 4-grams, i.e., sequence of 4 words. 2) ROUGE-L measures the harmonic mean of precision and recall on the longest common subsequence (LCS) between generated sentence and ground-truth sentence.  3) METEOR computes its score by utilizing a weighted F-score based on unigrams, and incorporating a penalty function to penalize the incorrect word order in the generated sentence. 4) CIDER utilizes a voting-based method to have a robust measurement against noise or incorrect annotations.

\subsection{Implementation Details}
\subsubsection{X-CLIP}
The feature extractor for both video features and text features is a pretrained CLIP (VIT-B/16) \cite{Radford2021LearningTV} model, which was pretrained on a large-scale image-text dataset. The learning rate in this experiment is set to 1e-4 after carefully tuning. For the maximum word length, maximum frame length, and the number of training epochs, we set the hyperparameters to 32, 12, and 5, respectively. We set the batch size for the training to 16, and apply the gradient accumulation technique to fit the batch of data into the GPU memory. Our experiments are conducted on a Linux environment computer with 1 NVIDIA GeForce GTX 1080 Ti, which takes about 15 hours for training on our dataset. The X-CLIP model is implemented using the PyTorch library.

\subsubsection{VNS-GRU}
We extract the video features using Efficient Convolutional Network (ECN) \cite{10.1007/978-3-030-01216-8_43} which was pretrained on Kinetics-400 dataset \cite{DBLP:journals/corr/KayCSZHVVGBNSZ17}. The features are extracted from the global pooling layers of the network with dimension 1536. The semantic features are the concatenation of the features extracted from the probabilities output of SCD \cite{SCN_CVPR2017} and ResNeXt-101 \cite{8100117}. For the text features, the Indonesian word vectors are extracted using fastText \cite{bojanowski2016enriching}, in which the model was trained using continuous bag-of-words (CBOW) with position-weights and dimension 300. The learning rate in this experiment is tuned and set to 3e-4. The number of sampled annotations is fixed to 4 for the training. We set the batch size and the training epochs to 128 and 50, respectively. For this captioning task, we conducted the experiments using 1 NVIDIA GeForce GTX 1650, which takes around 43 minutes for the training. The TensorFlow library is used to implement the VNS-GRU model.

\begin{table*}
\footnotesize
  \centering
  \begin{tabular}{l | l l l l l | l l l l l}
    \toprule
         & \multicolumn{5}{c |}{Text-to-Video Retrieval} & \multicolumn{5}{c}{Video-to-Text Retrieval} \\
    \midrule
    Method & R@1$\uparrow$ & R@5$\uparrow$ & R@10$\uparrow$ & MdR$\downarrow$ & MnR$\downarrow$ & R@1$\uparrow$ & R@5$\uparrow$ & R@10$\uparrow$ & MdR$\downarrow$ & MnR$\downarrow$ \\
    \midrule
    w/ Temporal Encoder & 32.2 & 62.9 & 74.3 & \textbf{3.0} & 17.8 & 39.9 & 71.6 & 82.6 & \textbf{2.0} & 11.6     \\
    w/o Temporal Encoder    & \textbf{32.3} & \textbf{63.3} & \textbf{74.9} & \textbf{3.0} & \textbf{17.5} & \textbf{44.9} & \textbf{77.6} & \textbf{88.8} & \textbf{2.0} & \textbf{6.4}      \\
    \bottomrule
  \end{tabular}
  \caption{{Impact of temporal encoder module in the X-CLIP algorithm on the MSVD-Indonesian dataset. The symbol $\uparrow$ indicates the higher value in the metric is better, while the symbol $\downarrow$ indicates the lower value in the metric is better.}}
  \label{tab:temporalencoder}
\end{table*}

\begin{table*}
 \small
  \centering
  \begin{tabular}{c c | l l l l l | l l l l l}
    \toprule
         \multicolumn{2}{c |}{Pretrained CLIP} & \multicolumn{5}{c |}{Text-to-Video Retrieval} & \multicolumn{5}{c}{Video-to-Text Retrieval} \\
    \midrule
    Visual & Text (EN) & R@1$\uparrow$ & R@5$\uparrow$ & R@10$\uparrow$ & MdR$\downarrow$ & MnR$\downarrow$ & R@1$\uparrow$ & R@5$\uparrow$ & R@10$\uparrow$ & MdR$\downarrow$ & MnR$\downarrow$ \\
    \midrule
    \xmark & \xmark & 0.8 & 2.3 & 4.4 & 199.0 & 234.5 & 0.5 & 2.4 & 5.1 & 189.0 & 224.6     \\
    \xmark & \cmark & 1.6 & 5.2 & 8.7 & 149.0 & 196.3 & 1.0 & 3.3 & 5.8 & 151.0 & 199.3     \\
    \cmark & \xmark & 12.7 & 34.7 & 47.7 & 12.0 & 53.2 & 9.2 & 33.5 & 48.1 & 11.0 & 37.0      \\
    \cmark & \cmark & \textbf{32.3} & \textbf{63.3} & \textbf{74.9} & \textbf{3.0} & \textbf{17.5} & \textbf{44.9} & \textbf{77.6} & \textbf{88.8} & \textbf{2.0} & \textbf{6.4}      \\
    \bottomrule
  \end{tabular}
  \caption{{Impact of pretrained CLIP in the X-CLIP algorithm on the MSVD-Indonesian dataset. The symbol $\uparrow$ indicates the higher value in the metric is better, while the symbol $\downarrow$ indicates the lower value in the metric is better. Initialization scheme for X-CLIP encoders:  \cmark indicates the encoder weights are initialized using the pretrained CLIP model, while \xmark\ indicates random initialization.}}
  \label{tab:pretrainedclip}
\end{table*}

\begin{table*}[ht!]
  \centering
  \begin{tabular}{l | l l l l l | l l l l l}
    \toprule
         & \multicolumn{5}{c |}{Text-to-Video Retrieval} & \multicolumn{5}{c}{Video-to-Text Retrieval} \\
    \midrule
    Model & R@1$\uparrow$ & R@5$\uparrow$ & R@10$\uparrow$ & MdR$\downarrow$ & MnR$\downarrow$ & R@1$\uparrow$ & R@5$\uparrow$ & R@10$\uparrow$ & MdR$\downarrow$ & MnR$\downarrow$ \\
    \midrule
    ViT-B/32 & 28.1 & 58.6 & 70.9 & 4.0 & 20.6 & 35.4 & 68.0 & 78.4 & 3.0 & 13.6     \\
    ViT-B/16    & \textbf{32.3} & \textbf{63.3} & \textbf{74.9} & \textbf{3.0} & \textbf{17.5} & \textbf{44.9} & \textbf{77.6} & \textbf{88.8} & \textbf{2.0} & \textbf{6.4}      \\
    \bottomrule
  \end{tabular}
  \caption{{Impact of different CLIP models in the X-CLIP algorithm on the MSVSD-Indonesian dataset. The symbol $\uparrow$ indicates the higher value in the metric is better, while the symbol $\downarrow$ indicates the lower value in the metric is better.}}
  \label{tab:vit}
\end{table*}

\subsection{Quantitative Results}
\subsubsection{X-CLIP}
\textbf{Is the temporal encoder module needed on a less complex dataset?}
Temporal encoder module \cite{10.1145/3474085.3479207} is a 3-layer transformer that is proposed in the X-CLIP architecture to capture temporal interaction between different frames. Although this module is expected to improve the accuracy of the model, the additional parameters introduced may result in sub-optimal performance on a small dataset \cite{LUO2022293}. As shown in Table~\ref{tab:temporalencoder}, we found that adding the temporal encoder also does not help to improve the performance of the model on our MSVD-Indonesian dataset. For most of the metrics, we can see that without using the temporal encoder module, the X-CLIP model can even outperform the one using the temporal encoder module with a decent margin. On the video-to-retrieval task, the performance gain is 5, 6, and 6.2 points in R@1, R@5, and R@10 metrics. On the text-to-video retrieval task, the performance gain is 0.1, 0.4, and 0.6 in R@1, R@5, and R@10 metrics. We expect these results due to the different characteristics of our dataset compared to the English MSVD dataset. In our dataset, the length of the sentences tends to be shorter, and the vocabulary size is comparatively smaller.

\textbf{Is CLIP model pretrained on English image-text dataset useful for our Indonesian video-text dataset?}
In Table~\ref{tab:pretrainedclip}, we investigate the impact of the CLIP model of the X-CLIP algorithm which was primarily pretrained on the English image-text dataset. When the petrained visual or text encoder of the CLIP model is not used to initialize the X-CLIP encoders, we initialize the X-CLIP encoders with random values. For the text encoder, if the pretrained weights from the CLIP model are not used, we replace the original CLIP tokenizer with the BERT tokenizer for the Indonesian language \cite{koto2020indolem}. From the table, we can see that incorporating the English pretrained CLIP model, both the visual and text encoder, can significantly help to improve performance. Although the text encoder is not specifically pretrained on the Indonesian language, the general linguistic pattern and semantic relationships learned in the pretrained CLIP model may still provide valuable information when it is applied to our MSVD-Indonesian dataset.

\begin{table}[ht]

  \centering
  \begin{tabular}{c| l l l l}
    \toprule
    SCD (En) & B4 & C & M & R \\
    \midrule
    \xmark & 38.30 & 84.48 & 32.02 & 65.07    \\
    \cmark & \textbf{58.68} & \textbf{126.65} & \textbf{40.33} & \textbf{76.84}     \\
    \bottomrule
  \end{tabular}
  \caption{{Ablation study of the SCD model in the VNS-GRU Algorithm on the MSVD-Indonesian dataset.}}
  \label{tab:scd}
\end{table}

\begin{table}[ht!]
  \centering
  \begin{tabular}{l | l l l l}
    \toprule
    Method & B4 & C & M & R \\
    \midrule
    2 & 54.96 & 121.43 & 39.22 & 75.85    \\
    4 & \textbf{58.68} & \textbf{126.65} & \textbf{40.33} & \textbf{76.84}    \\
    8 & 58.32 & 125.14 & 40.12 & 76.76    \\
    16 & 56.91 & 125.70 & 40.06 & 76.66    \\
    EXP & 56.89 & 122.87 & 39.79 & 76.64    \\
    \bottomrule
  \end{tabular}
  \caption{{Impact of sampling numbers annotations in the VNS-GRU algorithm on the MSVD-Indonesian dataset. EXP denotes a non-fixed sampling schedule, i.e., exponential schedule, as defined in equation (25) in \cite{DBLP:conf/ecai/Chen0020a}.}}
  \label{tab:samplingnumber}
\end{table}

\textbf{How do the different CLIP models affect the results?}
We further investigate different CLIP models, i.e., ViT-B/16 and ViT-B/32, on our dataset as presented in Table~\ref{tab:vit}. The results demonstrate that the X-CLIP model utilizing ViT-B/16 consistently outperforms the model utilizing ViT-B/32 across all evaluation metrics on the MSVD-Indonesian dataset. These findings are consistent with the experiments conducted in \cite{10.1145/3503161.3547910}, where the X-CLIP model utilizing ViT-B/16 exhibited superior performance compared to the one utilizing ViT-B/32 on the MSVD dataset.

\subsubsection{VNS-GRU}

\textbf{Is SCD model pretrained on English video-text dataset useful for our Indonesian video-text dataset?} We investigate the usage of pretrained SCD as a feature extractor for the VNS-GRU model in Table~\ref{tab:scd}. The SCD model was pretrained on the English MSVD dataset to investigate the cross-lingual knowledge transfer from English to our Indonesian dataset. From the table, we can see that the performance gain is 20.38, 42.17, 8.31, and 11.77 points in BLEU@4, CIDEr, METEOR, and ROUGE-L metrics, respectively. These results indicate that the pretrained SCD model on the English video-text dataset can be employed to extract useful semantic information, which can be transferred to the Indonesian language. We expect this because most of the top $n$ vocabulary extracted in the English dataset, which was used to pretrained the SCD model, is still semantically similar to the ones in our MSVD-Indonesian dataset. Although the performance may be improved further by using the SCD model which specifically pretrained on the Indonesian video-text dataset, the investigation for those works is left for future study.

\textbf{How do the different sampling numbers of annotations affect the results?} In the VNS-GRU algorithm, the training phase is divided into two phases. The first phase is all the annotations are equally used during training. In the second phase, a number of annotations are sampled, with the motivation to avoid only focusing on common words and forgetting detailed words. In this experiment, as shown in Table~\ref{tab:samplingnumber}, we conduct an ablation study of different sampling numbers of annotations on the configuration of the VNS-GRU algorithm. Chen \etal showed that a fixed sample size of 16 is the best configuration on the MSVD dataset. In our experiment, we found that using 4 as the sampling number is the best on our MSVD-Indonesian dataset. We expect this behavior because in our dataset, the average length of the sentences is relatively shorter and the vocabulary size is relatively smaller. When the sentences are simpler and have more words in common, a model is able to achieve better performance by focusing on fewer sentences \cite{DBLP:conf/ecai/Chen0020a}.

\begin{figure*}[htbp]
  \begin{subfigure}[b]{\textwidth}
    \centering
    \includegraphics[width=\textwidth]{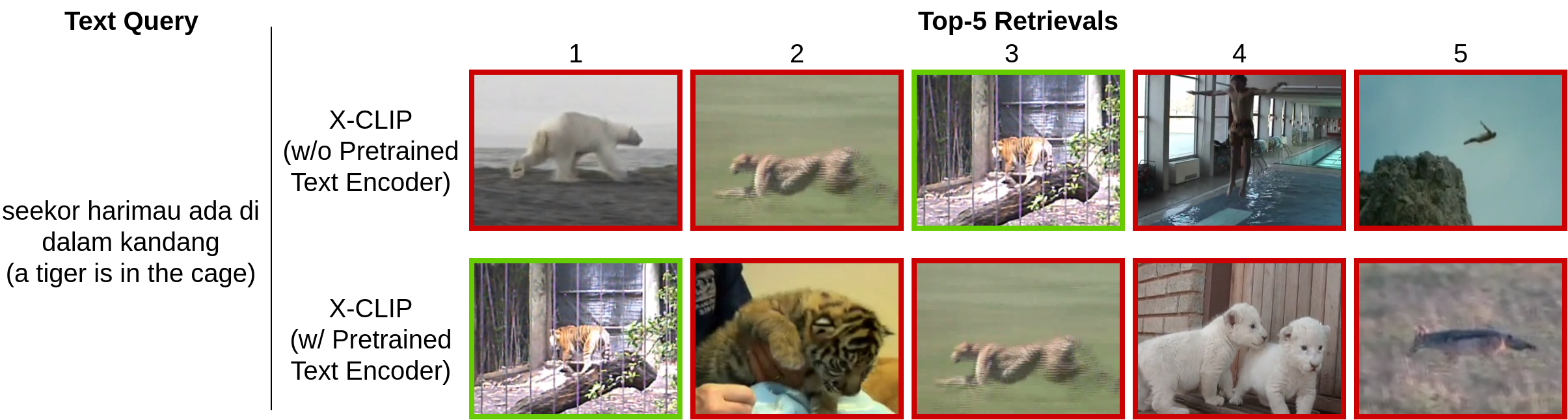}
    \caption{}
    \label{fig:t2v1}
  \end{subfigure}
  \vfill
  \begin{subfigure}[b]{\textwidth}
    \centering
    \includegraphics[width=\textwidth]{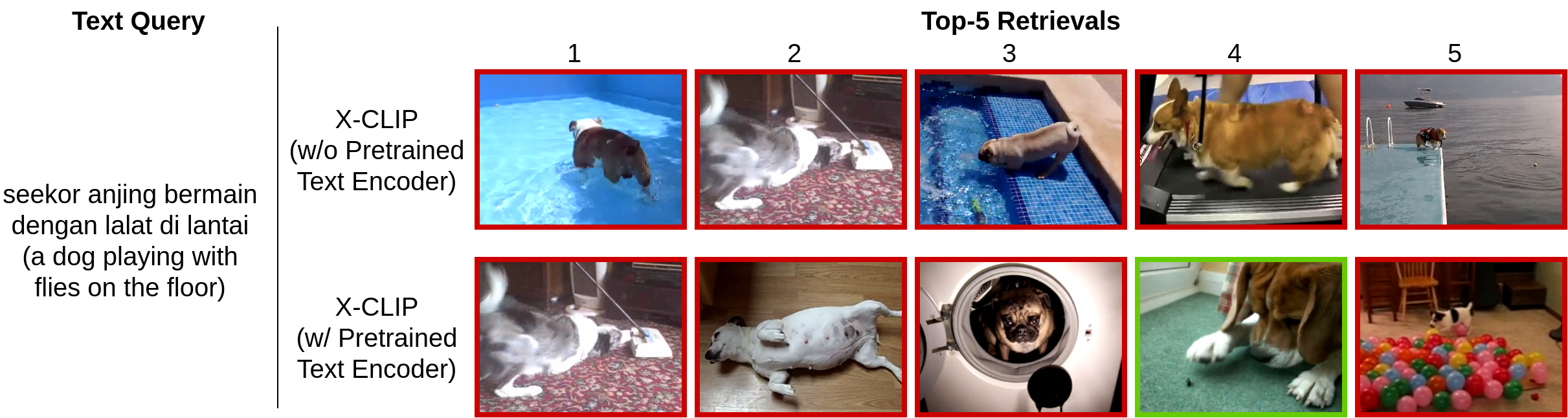}
    \caption{}
    \label{fig:t2v2}
  \end{subfigure}
  
  \caption{Qualitative results of the X-CLIP model without vs with the pretrained text encoder of CLIP model on the text-to-video retrieval task. When the pretrained text encoder of the CLIP model is not used, the text encoder weights of the X-CLIP model are randomly initialized. With respect to the ground truth, the green box and the red box indicate the relevant and the irrelevant video, respectively.}
  \label{fig:qual-t2v}
\end{figure*}

\begin{figure*}[t!]
  \centering  
  \begin{subfigure}[b]{\textwidth}
    \centering
    \includegraphics[width=\textwidth]{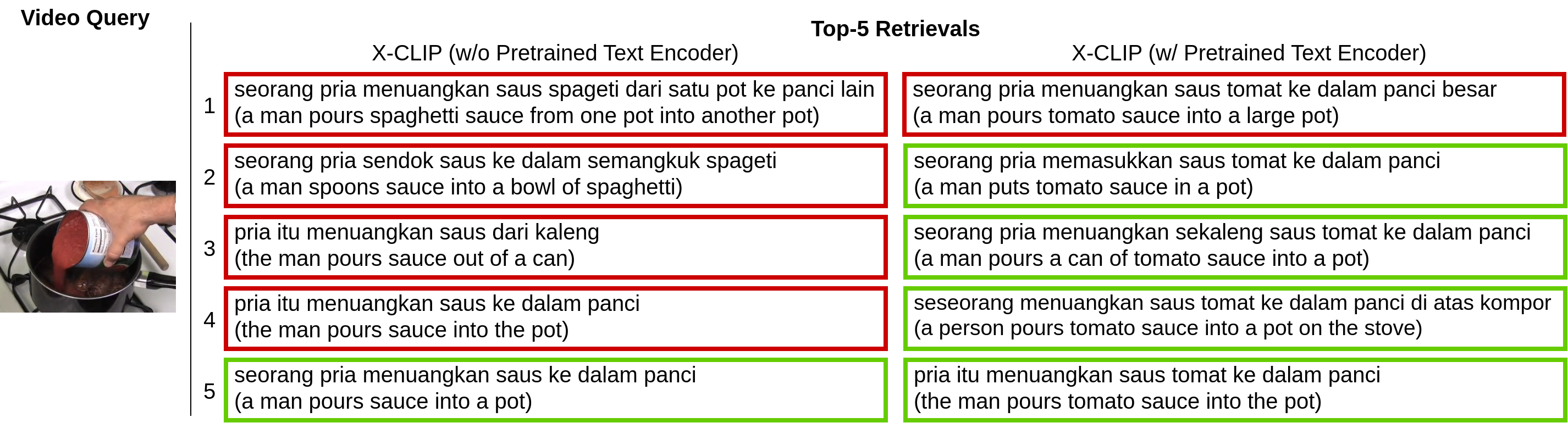}
    \caption{}
    \label{fig:v2t1}
  \end{subfigure}
  \vfill
  \begin{subfigure}[b]{\textwidth}
    \centering
    \includegraphics[width=\textwidth]{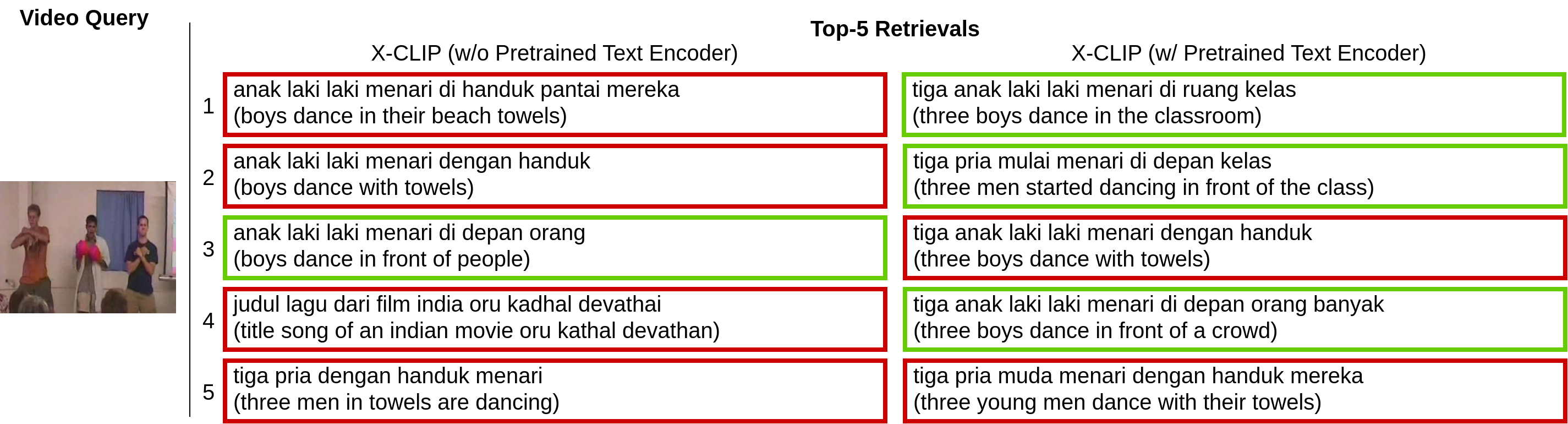}
    \caption{}
    \label{fig:v2t2}
  \end{subfigure}
  
  \caption{Qualitative results of the X-CLIP model without vs with the pretrained text encoder of CLIP model on the video-to-text retrieval task. When the pretrained text encoder of the CLIP model is not used, the text encoder weights of the X-CLIP model are randomly initialized. With respect to the ground truth, the green box and the red box indicate the relevant and the irrelevant text, respectively.}
  \label{fig:qual-v2t}
\end{figure*}

\subsection{Qualitative Results}
We show qualitative results of our experiments on the MSVD-Indonesian dataset for text-to-video retrieval, video-to-text retrieval, and video captioning in Figure~\ref{fig:qual-t2v}, Figure~\ref{fig:qual-v2t}, and Figure~\ref{fig:qual-cap}, respectively. For the X-CLIP model, i.e., the retrieval model, we compare the top-5 retrieval results obtained between 1) the text encoder is randomly initialized, i.e., without using the pretrained text encoder of the CLIP model, and 2) the text encoder is initialized from the pretrained CLIP model. In both cases, the weights of the visual encoder are initialized with the pretrained visual encoder of the CLIP model. For the VNS-GRU model, i.e., captioning model, we compare the results obtained between 1) pretrained SCD model is not used and 2) the pretrained SCD model is used.

\subsubsection{X-CLIP}
In the text-to-video retrieval results, as illustrated in Figure~\ref{fig:qual-t2v}, we observe that the X-CLIP model with random initialization on the text encoder does not retrieve the relevant video as accurately as the model with the text encoder weights initialized from the CLIP model. As shown in Figure~\ref{fig:qual-t2v} (a), X-CLIP with random initialization in the text encoder is still able to retrieve the relevant video (green) at the 3rd rank. However, incorporating the pretrained initialization on both visual and text encoder helps to further improve the retrieval results by having the relevant video placed at the 1st rank and discarding the highly irrelevant video from the top-5 retrievals, i.e., videos containing 'human' in the 4th and 5th rank in the figure. Although the replacement at the 4th and the 5th rank is still not exactly relevant to the given text query, retrieving the videos of animals are arguably more relevant than retrieving videos of humans given the query which includes 'tiger' in the sentence. In Figure~\ref{fig:qual-t2v} (b), we can observe that the X-CLIP model with the pretrained text encoder successfully retrieves the relevant video in the top-5 retrieval results. Conversely, the model without the pretrained text encoder, i.e., random initialization, fails to retrieve the relevant video.

In the video-to-text results as shown in Figure~\ref{fig:qual-v2t}, we can notice that the number of relevant texts, w.r.t. the ground truth, in the top-5 retrievals are different between the two X-CLIP models with different text encoder initialization. In Figure~\ref{fig:qual-v2t} (a), the X-CLIP model with random initialization on the text encoder is still able to retrieve the relevant text at the 5th rank. The X-CLIP model with the pretrained text encoder, however, is able to retrieve more relevant texts in the top-5 results. Interestingly, except for the retrieved text in the 2nd rank of results obtained from the model without the pretrained text encoder, the other irrelevant texts (red) in the figure are still semantically well aligned to the video query. Similarly, as shown in Figure~\ref{fig:qual-v2t} (b), we also observe that the X-CLIP model with the randomly initialized text encoder is still capable of retrieving relevant text. However, the model with the pretrained text encoder performs better, yielding more relevant texts in the top-5 retrieval results.

Furthermore, while our experimental results demonstrate the capabilities of the X-CLIP model in retrieving relevant videos and texts on our MSVD-Indonesian dataset, it is clear that there is still room for improvement. Figure~\ref{fig:qual-t2v} and Figure~\ref{fig:qual-v2t} show instances where irrelevant videos and texts are retrieved in the top-5 retrieval results. Future research efforts can focus on enhancing the relevance of retrieved videos and texts, resulting in more precise and comprehensive results.

\subsubsection{VNS-GRU}
\begin{figure*}[htbp]
  \centering
  \begin{subfigure}[b]{0.495\textwidth}
    \centering
    \includegraphics[width=\textwidth]{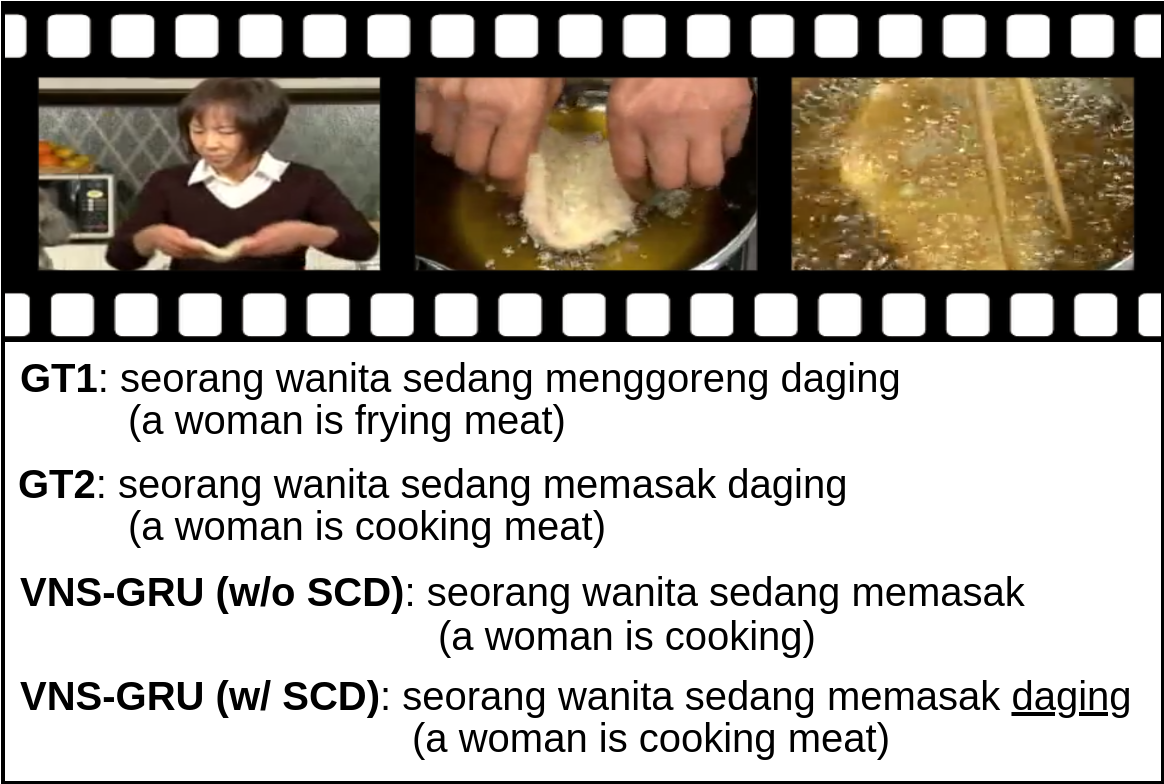}
    \caption{}
    \label{fig:cap1}
  \end{subfigure}
  \hfill
  \begin{subfigure}[b]{0.495\textwidth}
    \centering
    \includegraphics[width=\textwidth]{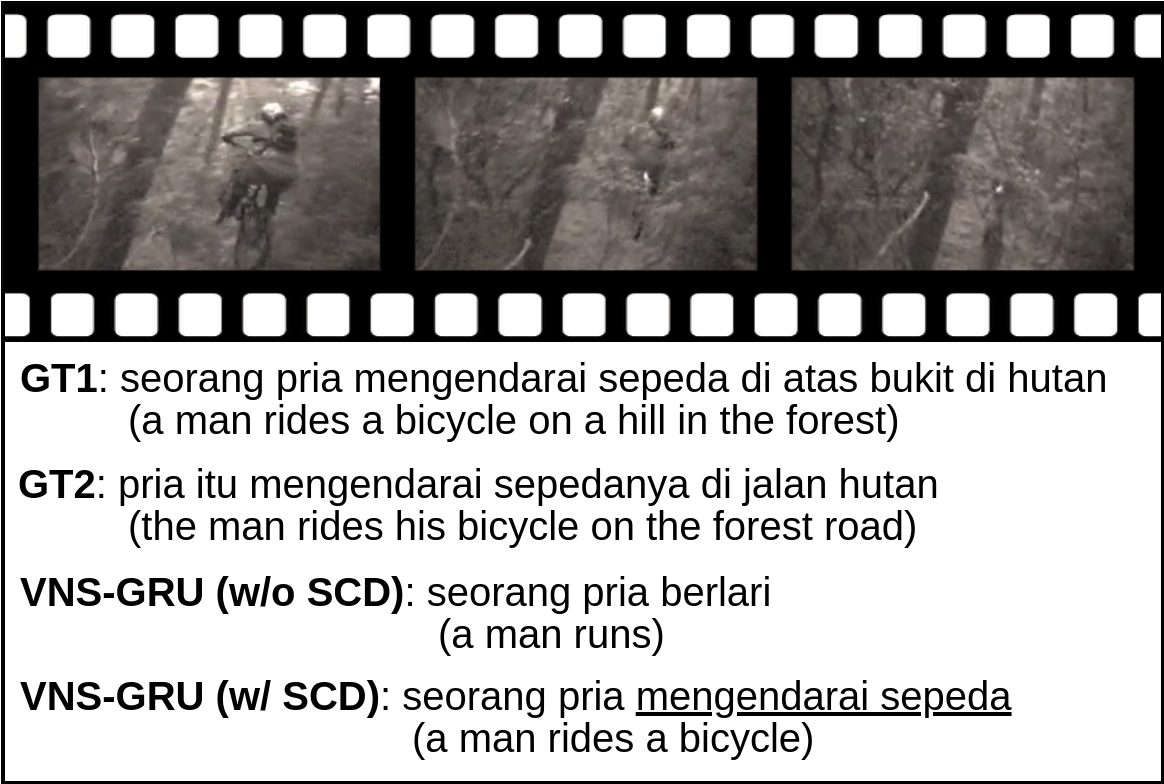}
    \caption{}
    \label{fig:cap2}
  \end{subfigure}
  
  \caption{Qualitative results of the VNS-GRU model without vs with SCD model on the video captioning task.}
  \label{fig:qual-cap}
\end{figure*}

In Figure~\ref{fig:qual-cap}, we can see that the VNS-GRU model without the SCD model pretrained on the English video-text dataset does not generate sentences better compared to the one which employs the pretrained SCD model. From Figure~\ref{fig:qual-cap} (a), we observe that the utilization of the English SCD model helps to generate a more details sentence with the word `daging (meat)' included in the sentence. From Figure~\ref{fig:qual-cap} (b), we can notice that the absence of the SCD model may generate an inaccurate sentence. Incorporating the pretrained SCD model guides the model to better capture the action and the object in the video, i.e., `mengendarai (rides)' and 'sepeda (bicycle)'. Although the SCD model was pretrained on the English video-text dataset, these results show that the extracted semantic information from the model could still be useful for training the VNS-GRU model on our MSVD-Indonesian dataset. 

Despite our experimental results indicating that the model is able to generate Indonesian sentences with reasonable accuracy, one can see that the generated sentences still lack sufficient details. This suggests that there is potential for further improvement in capturing specific details in the generated text. This can involve exploring techniques to incorporate more contextual information, improving the modeling of fine-grained details, and refining the language generation process to produce more informative and detailed sentences.

\section{Discussion, Future Works and Conclusion}
\label{sec:future}
In this work, we constructed the MSVD-Indonesian dataset, which is the first public video-text dataset in the Indonesian language at the time of writing this paper. We conducted experiments and discussed the performance of two neural network models on our dataset for three different tasks, i.e., text-to-video retrieval, video-to-text retrieval, and video captioning. Our experimental results showed that the prior work which performed well on the English video-text dataset, i.e., the MSVD dataset, could also be applicable to our MSVD-Indonesian dataset with some modifications on the algorithms or on the parameters. Besides, incorporating a feature extractor that was pretrained on the English vision-language dataset could also help to improve the performance of the models on our Indonesian video-text dataset. Through our study, we also found that there are several potential works that can be explored for future research.

\textbf{Pretraining on a large-scale vision-language dataset.} A neural network model pretrained on a large-scale vision-language dataset has been widely adopted in many recent algorithms on many research tasks. The pretrained model serves as a powerful feature extractor that can boost the accuracy significantly. In our experiments, we assumed that there is a lack of pretraining resources for the Indonesian language. To address the issue, we utilized the models which mainly pretrained on the English vision-language dataset as the feature extractors. Future studies could delve into the exploration of pretraining the models on a large-scale Indonesian vision-language dataset, as this has the potential to substantially enhance the accuracy of the final models.

\textbf{Multilingual Output.} We conducted experiments on several research tasks focusing on monolingual output, where only the Indonesian language is outputted by a model. However, it is also interesting to develop an algorithm which able to output sentences in multiple languages given a video as the input. Since each Indonesian sentence in our dataset has a corresponding English sentence in the original MSVD dataset, exploring a multilingual approach becomes an interesting prospect for future work.

\textbf{Noise-Robust Algorithm.} We have shown that our dataset includes some inaccurate sentences due to the limitation of the machine translation service and also the inherited inaccurate annotations from the English dataset. Our experiments showed that the models are still able to produce output reasonably well despite the inaccuracy in the sentences. Yet, the existence of the noises is not explicitly addressed and investigated yet.  Investigating the impact of the noise and developing a noise-robust algorithm may also be interesting research to be explored.

In conclusion, we believe that our MSVD-Indonesian dataset can be used as an important benchmark for multiple video-text tasks, including text-to-video retrieval, video-to-text retrieval, and video captioning. Our benchmark dataset can encourage innovation in building a better algorithm for multimodal video-text research in the Indonesian language.
{
    \small
    \bibliographystyle{ieeenat_fullname}
    \bibliography{main}

\begin{thebibliography}{30}
\providecommand{\natexlab}[1]{#1}
\providecommand{\url}[1]{\texttt{#1}}
\expandafter\ifx\csname urlstyle\endcsname\relax
  \providecommand{\doi}[1]{doi: #1}\else
  \providecommand{\doi}{doi: \begingroup \urlstyle{rm}\Url}\fi

\bibitem[Baltrusaitis et~al.(2019)Baltrusaitis, Ahuja, and Morency]{10.1109/TPAMI.2018.2798607}
Tadas Baltrusaitis, Chaitanya Ahuja, and Louis-Philippe Morency.
\newblock Multimodal machine learning: A survey and taxonomy.
\newblock \emph{IEEE Trans. Pattern Anal. Mach. Intell.}, 41\penalty0 (2):\penalty0 423–443, 2019.

\bibitem[Banerjee and Lavie(2005)]{banarjee2005}
Satanjeev Banerjee and Alon Lavie.
\newblock {METEOR}: An automatic metric for {MT} evaluation with improved correlation with human judgments.
\newblock In \emph{Proceedings of the {ACL} Workshop on Intrinsic and Extrinsic Evaluation Measures for Machine Translation and/or Summarization}, pages 65--72, Ann Arbor, Michigan, 2005. Association for Computational Linguistics.

\bibitem[Bojanowski et~al.(2016)Bojanowski, Grave, Joulin, and Mikolov]{bojanowski2016enriching}
Piotr Bojanowski, Edouard Grave, Armand Joulin, and Tomas Mikolov.
\newblock Enriching word vectors with subword information.
\newblock \emph{arXiv preprint arXiv:1607.04606}, 2016.

\bibitem[Chen and Dolan(2011)]{chen:acl11}
David~L. Chen and William~B. Dolan.
\newblock Collecting highly parallel data for paraphrase evaluation.
\newblock In \emph{Proceedings of the 49th Annual Meeting of the Association for Computational Linguistics (ACL-2011)}, Portland, OR, 2011.

\bibitem[Chen et~al.(2020{\natexlab{a}})Chen, Li, and Hu]{DBLP:conf/ecai/Chen0020a}
Haoran Chen, Jianmin Li, and Xiaolin Hu.
\newblock Delving deeper into the decoder for video captioning.
\newblock In \emph{{ECAI} 2020 - 24th European Conference on Artificial Intelligence, 29 August-8 September 2020, Santiago de Compostela, Spain, August 29 - September 8, 2020 - Including 10th Conference on Prestigious Applications of Artificial Intelligence {(PAIS} 2020)}, pages 1079--1086. {IOS} Press, 2020{\natexlab{a}}.

\bibitem[Chen et~al.(2020{\natexlab{b}})Chen, Lin, Maye, Li, and Hu]{10.3389/frobt.2020.475767}
Haoran Chen, Ke Lin, Alexander Maye, Jianmin Li, and Xiaolin Hu.
\newblock A semantics-assisted video captioning model trained with scheduled sampling.
\newblock \emph{Frontiers in Robotics and AI}, 7:\penalty0 129, 2020{\natexlab{b}}.

\bibitem[Gan et~al.(2017)Gan, Gan, He, Pu, Tran, Gao, Carin, and Deng]{SCN_CVPR2017}
Zhe Gan, Chuang Gan, Xiaodong He, Yunchen Pu, Kenneth Tran, Jianfeng Gao, Lawrence Carin, and Li Deng.
\newblock Semantic compositional networks for visual captioning.
\newblock In \emph{CVPR}, 2017.

\bibitem[Kay et~al.(2017)Kay, Carreira, Simonyan, Zhang, Hillier, Vijayanarasimhan, Viola, Green, Back, Natsev, Suleyman, and Zisserman]{DBLP:journals/corr/KayCSZHVVGBNSZ17}
Will Kay, Jo{\~{a}}o Carreira, Karen Simonyan, Brian Zhang, Chloe Hillier, Sudheendra Vijayanarasimhan, Fabio Viola, Tim Green, Trevor Back, Paul Natsev, Mustafa Suleyman, and Andrew Zisserman.
\newblock The kinetics human action video dataset.
\newblock \emph{CoRR}, abs/1705.06950, 2017.

\bibitem[Koto et~al.(2020)Koto, Rahimi, Lau, and Baldwin]{koto2020indolem}
Fajri Koto, Afshin Rahimi, Jey~Han Lau, and Timothy Baldwin.
\newblock Indolem and indobert: A benchmark dataset and pre-trained language model for indonesian nlp.
\newblock In \emph{Proceedings of the 28th COLING}, 2020.

\bibitem[Krishna et~al.(2017)Krishna, Hata, Ren, Fei-Fei, and Niebles]{krishna2017dense}
Ranjay Krishna, Kenji Hata, Frederic Ren, Li Fei-Fei, and Juan~Carlos Niebles.
\newblock Dense-captioning events in videos.
\newblock In \emph{International Conference on Computer Vision (ICCV)}, 2017.

\bibitem[Lin(2004)]{lin2004rouge}
Chin-Yew Lin.
\newblock Rouge: a package for automatic evaluation of summaries.
\newblock In \emph{Workshop on Text Summarization Branches Out, Post-Conference Workshop of ACL 2004, Barcelona, Spain}, pages 74--81, 2004.

\bibitem[Luo et~al.(2022)Luo, Ji, Zhong, Chen, Lei, Duan, and Li]{LUO2022293}
Huaishao Luo, Lei Ji, Ming Zhong, Yang Chen, Wen Lei, Nan Duan, and Tianrui Li.
\newblock Clip4clip: An empirical study of clip for end to end video clip retrieval and captioning.
\newblock \emph{Neurocomputing}, 508:\penalty0 293--304, 2022.

\bibitem[Ma et~al.(2022)Ma, Xu, Sun, Yan, Zhang, and Ji]{10.1145/3503161.3547910}
Yiwei Ma, Guohai Xu, Xiaoshuai Sun, Ming Yan, Ji Zhang, and Rongrong Ji.
\newblock X-clip: End-to-end multi-grained contrastive learning for video-text retrieval.
\newblock In \emph{Proceedings of the 30th ACM International Conference on Multimedia}, page 638–647, New York, NY, USA, 2022. Association for Computing Machinery.

\bibitem[{Media Computing and Intelligent Systems Lab - Beijing Institute of Technology}(2018)]{msvdcn}
{Media Computing and Intelligent Systems Lab - Beijing Institute of Technology}.
\newblock {MSVD-CN}.
\newblock \url{https://github.com/mcislab-machine-learning/MSVD-CN}, 2018.

\bibitem[Papineni et~al.(2002)Papineni, Roukos, Ward, and Zhu]{10.3115/1073083.1073135}
Kishore Papineni, Salim Roukos, Todd Ward, and Wei-Jing Zhu.
\newblock Bleu: A method for automatic evaluation of machine translation.
\newblock In \emph{Proceedings of the 40th Annual Meeting on Association for Computational Linguistics}, page 311–318, USA, 2002. Association for Computational Linguistics.

\bibitem[Perez-Martin et~al.(2021{\natexlab{a}})Perez-Martin, Bustos, Guimar{\~a}es, Sipiran, P'erez, and Said]{PerezMartin2021ACR}
Jesus Perez-Martin, Benjam{\'i}n Bustos, Silvio Jamil~Ferzoli Guimar{\~a}es, Ivan Sipiran, Jorge~A. P'erez, and Grethel~Coello Said.
\newblock A comprehensive review of the video-to-text problem.
\newblock \emph{Artificial Intelligence Review}, 55:\penalty0 4165 -- 4239, 2021{\natexlab{a}}.

\bibitem[Perez-Martin et~al.(2021{\natexlab{b}})Perez-Martin, Bustos, and Perez]{Perez-Martin_2021_WACV}
Jesus Perez-Martin, Benjamin Bustos, and Jorge Perez.
\newblock Improving video captioning with temporal composition of a visual-syntactic embedding.
\newblock In \emph{Proceedings of the IEEE/CVF Winter Conference on Applications of Computer Vision (WACV)}, pages 3039--3049, 2021{\natexlab{b}}.

\bibitem[Perez-Martin et~al.(2021{\natexlab{c}})Perez-Martin, Bustos, and Pérez]{9412898}
Jesus Perez-Martin, Benjamin Bustos, and Jorge Pérez.
\newblock Attentive visual semantic specialized network for video captioning.
\newblock In \emph{2020 25th International Conference on Pattern Recognition (ICPR)}, pages 5767--5774, 2021{\natexlab{c}}.

\bibitem[Radford et~al.(2021)Radford, Kim, Hallacy, Ramesh, Goh, Agarwal, Sastry, Askell, Mishkin, Clark, Krueger, and Sutskever]{Radford2021LearningTV}
Alec Radford, Jong~Wook Kim, Chris Hallacy, Aditya Ramesh, Gabriel Goh, Sandhini Agarwal, Girish Sastry, Amanda Askell, Pamela Mishkin, Jack Clark, Gretchen Krueger, and Ilya Sutskever.
\newblock Learning transferable visual models from natural language supervision.
\newblock In \emph{International Conference on Machine Learning}, 2021.

\bibitem[Scaiella et~al.(2019)Scaiella, Croce, and Basili]{IJCOL:scaiella_et_al:2019}
Antonio Scaiella, Danilo Croce, and Roberto Basili.
\newblock Large scale datasets for image and video captioning in italian.
\newblock \emph{Italian Journal of Computational Linguistics}, 2\penalty0 (5):\penalty0 49--60, 2019.

\bibitem[Singh et~al.(2022)Singh, Singh, and Bandyopadhyay]{10.1007/s00530-021-00816-3}
Alok Singh, Thoudam~Doren Singh, and Sivaji Bandyopadhyay.
\newblock Attention based video captioning framework for hindi.
\newblock \emph{Multimedia Syst.}, 28\penalty0 (1):\penalty0 195–207, 2022.

\bibitem[Tang et~al.(2021)Tang, Wang, LIU, Rao, Li, and Li]{10.1145/3474085.3479207}
Mingkang Tang, Zhanyu Wang, Zhenhua LIU, Fengyun Rao, Dian Li, and Xiu Li.
\newblock Clip4caption: Clip for video caption.
\newblock In \emph{Proceedings of the 29th ACM International Conference on Multimedia}, page 4858–4862, New York, NY, USA, 2021. Association for Computing Machinery.

\bibitem[Torabi et~al.(2016)Torabi, Tandon, and Sigal]{DBLP:journals/corr/TorabiTS16}
Atousa Torabi, Niket Tandon, and Leonid Sigal.
\newblock Learning language-visual embedding for movie understanding with natural-language.
\newblock \emph{CoRR}, abs/1609.08124, 2016.

\bibitem[Vedantam et~al.(2015)Vedantam, Zitnick, and Parikh]{7299087}
Ramakrishna Vedantam, C.~Lawrence Zitnick, and Devi Parikh.
\newblock Cider: Consensus-based image description evaluation.
\newblock In \emph{2015 IEEE Conference on Computer Vision and Pattern Recognition (CVPR)}, pages 4566--4575, 2015.

\bibitem[Venugopalan et~al.(2015)Venugopalan, Rohrbach, Donahue, Mooney, Darrell, and Saenko]{venugopalan15iccv}
Subhashini Venugopalan, Marcus Rohrbach, Jeff Donahue, Raymond Mooney, Trevor Darrell, and Kate Saenko.
\newblock Sequence to sequence -- video to text.
\newblock In \emph{Proceedings of the IEEE International Conference on Computer Vision (ICCV)}, 2015.

\bibitem[Wang et~al.(2019)Wang, Wu, Chen, Li, Wang, and Wang]{Wang_2019_ICCV}
Xin Wang, Jiawei Wu, Junkun Chen, Lei Li, Yuan-Fang Wang, and William~Yang Wang.
\newblock Vatex: A large-scale, high-quality multilingual dataset for video-and-language research.
\newblock In \emph{The IEEE International Conference on Computer Vision (ICCV)}, 2019.

\bibitem[Xie et~al.(2017)Xie, Girshick, Dollár, Tu, and He]{8100117}
Saining Xie, Ross Girshick, Piotr Dollár, Zhuowen Tu, and Kaiming He.
\newblock Aggregated residual transformations for deep neural networks.
\newblock In \emph{2017 IEEE Conference on Computer Vision and Pattern Recognition (CVPR)}, pages 5987--5995, 2017.

\bibitem[Xu et~al.(2016)Xu, Mei, Yao, and Rui]{xu2016msr-vtt}
Jun Xu, Tao Mei, Ting Yao, and Yong Rui.
\newblock Msr-vtt: A large video description dataset for bridging video and language.
\newblock IEEE International Conference on Computer Vision and Pattern Recognition (CVPR), 2016.

\bibitem[Zolfaghari et~al.(2018)Zolfaghari, Singh, and Brox]{10.1007/978-3-030-01216-8_43}
Mohammadreza Zolfaghari, Kamaljeet Singh, and Thomas Brox.
\newblock Eco: Efficient convolutional network for online video understanding.
\newblock In \emph{Computer Vision -- ECCV 2018}, pages 713--730, Cham, 2018. Springer International Publishing.

\bibitem[Çtamak et~al.(2019)Çtamak, Kuyu, Erdem, and Erdem]{8806555}
Begüm Çtamak, Menekşe Kuyu, Aykut Erdem, and Erkut Erdem.
\newblock Msvd-turkish: A large-scale dataset for video captioning in turkish.
\newblock In \emph{2019 27th Signal Processing and Communications Applications Conference (SIU)}, pages 1--4, 2019.

\end{thebibliography}
}

\end{document}